\documentclass[twocolumn,secnumarabic,amssymb,nobibnotes,aps,prd]{revtex4-2}

\usepackage{amssymb,amsmath}
\usepackage[dvipsnames]{xcolor}
\usepackage{hyperref}

\newcommand{\pd}{\partial}

\DeclareMathOperator{\sign}{sign}

\begin{document}

\title{Homogeneous Anisotropic Black Branes with Bianchi VI$_h$ Symmetry}
\author{Markus A. G. Amano}
\date{February 20, 2026}
\affiliation{Department of Electrical and Electronic Engineering, National Institute of Technology (NIT) Oyama College, Oyama, Japan}
\thanks{Prepared for submission to Phys. Rev. D.}

\begin{abstract}
We construct a new family of exact vacuum black brane solutions to five-dimensional Einstein gravity with a negative cosmological constant, characterized by a homogeneous horizon with Bianchi VI$_h$ symmetry.
This construction generalizes the known Solv (Bianchi VI$_{-1}$) geometry via a continuous anisotropy parameter, $h$.
By reducing the field equations to a cohomogeneity-one system, we derive the metric analytically.
These homogeneous solutions are not asymptotically (locally) AdS, but nevertheless we analyze the thermodynamics, establishing scaling relations between entropy and temperature with anisotropic holographic system in mind.
Additionally, we identify a new branch of Ricci-flat, hyperscaling-violating vacuum solutions in the case where the cosmological constant vanishes.
\end{abstract}

\maketitle

\tableofcontents

\section{Introduction}

Black holes and black branes with nontrivial horizon geometries arise naturally once one relaxes asymptotic flatness and/or considers higher-dimensional spacetimes.
In five bulk dimensions, spatial horizon sections are three-manifolds, and by Thurston's geometrization program any compact three-manifold admits a canonical decomposition into pieces modeled on the eight Thurston geometries \cite{Thurston:1997thr}.
Among these, the Nil, Solv, and $\widetilde{\mathrm{SL}}_2(\mathbb{R})$ geometries are homogeneous but neither of constant curvature nor simple products, making them natural laboratories for anisotropic gravitational dynamics.
Homogeneous horizon geometries of this type continue to play important roles in higher-dimensional gravity and black hole classification \cite{Bizon:2004wp,Kunduri:2007vf,Figueras:2011gd,Mann:1996gj,Lemos:1994xp,Iizuka:2012iv,Hervik:2003vx}.

A concrete realization of homogeneous Thurston horizons in asymptotically AdS gravity was given by Cadeau and Woolgar \cite{Cadeau:2000tj}, who constructed five-dimensional Einstein black holes whose horizons carry Nil and Solv geometries.
A key structural feature of these solutions is that when the spatial slices are written in a left-invariant coframe, the Einstein equations reduce to a cohomogeneity-one system of ordinary differential equations.
Motivated by these geometric ideas, Hassa\"ine and collaborators developed a systematic analysis of homogeneous anisotropic black branes and scaling geometries, including families with hyperscaling violation and horizons modeled on Thurston manifolds \cite{Hassaine:2015ifa,Bravo-Gaete:2017nkp}.
Related homogeneous holographic phases---such as helical (Bianchi VII$_0$) black branes, Q-lattices, and other homogeneous lattice constructions---preserve enough symmetry to keep the bulk problem effectively one-dimensional while still capturing anisotropic thermodynamics and transport \cite{Donos:2011ff,Donos:2013eha,Andrade:2013gsa,Donos:2014uba,Donos:2014oha,Alberte:2016xja,Amoretti:2017frz,Arean:2023tja}.
From a scaling perspective, many such solutions can be viewed as generalizations of Lifshitz and hyperscaling-violating geometries \cite{Kachru:2008yh,Charmousis:2010zz,Hendi:2015cra}, and systematic studies of Bianchi attractors clarify how different Bianchi types encode distinct patterns of anisotropy and symmetry breaking \cite{Iizuka:2012iv,Iizuka:2012pn,Cremonini:2012ir,Faedo:2022hle,Ren:2016xhb}.

In this paper, we construct a new vacuum solution of the five-dimensional Einstein equations with a negative cosmological constant whose constant-$r$ spatial sections are homogeneous of Bianchi VI$_h$ type, in a similar fashion to \cite{Cadeau:2000tj,Hassaine:2015ifa}, which model 3-hyperbolic space, $H^3$.
The Bianchi VI$_h$ algebra is a one-parameter family of solvable Lie algebras parameterized by $h\in\mathbb R$.
While special points in this family have appeared previously, a general Bianchi VI$_h$ vacuum black brane solution has not, to our knowledge, been exhibited outside cosmological applications \cite{Wainwright:1997,Feinstein:1993qm,Ellis:1990wsa}, non-vacuum solutions \cite{Donos:2013woa,Donos:2012js}, and nonconformal holography \cite{Giataganas:2017koz,Giataganas:2025ing,Donos:2016zpf,Iizuka:2012wt}.

An important structural feature of our solution is that although it solves the pure Einstein equations, it is neither asymptotically nor locally AdS.
Instead, the spacetime exhibits anisotropic power-law warping controlled by the parameter $h$, which suggests interpreting the geometry as an infrared scaling solution and, plausibly, as a near-horizon limit of a more general asymptotically locally AdS spacetime \cite{Mateos:2011tv,Mateos:2011ix,Cong:2024pvs,Liu:2014dva,Iizuka:2012wt}.
Indeed, in special limits (such as $h=0$ and $h=1$) the solution reduces to direct products involving AdS$_2$ or hyperbolic space, reinforcing the near-horizon/IR interpretation and motivating the question of whether there exist domain-wall solutions interpolating between AdS$_5$ in the UV and the Bianchi VI$_h$ geometry in the IR \cite{Baggioli:2024vza,Kachru:2013voa}.

The organization of the paper is as follows.
In Section~\ref{sec:theory} we present the Einstein setup, review the Bianchi VI$_h$ left-invariant one-forms, and introduce a cohomogeneity-one metric ansatz.
Section~\ref{sec:solution} contains the resulting family of black branes together with horizon regularity and asymptotic analysis, including special cases where $h=1$.
In Section~\ref{sec:thermo} we compute the basic thermodynamic quantities.
These geometries exhibit distinct power-law scaling behaviors and thermodynamic properties, extending the solution space to the strictly vacuum regime.
Finally, in Section~\ref{sec:lambda_zero} we consider the vanishing cosmological constant limit ($\Lambda \to 0$). While the primary focus of holographic studies is often asymptotically AdS spacetimes, we identify a new branch of Ricci-flat vacuum solutions within the Bianchi VI$_h$ class similar to the Nil solution found in \cite{Hassaine:2015ifa}.
We conclude in Section~\ref{sec:conclusion} with a summary and outlook.

\section{Theory}
\label{sec:theory}

\subsection{Action and equations of motion}
We consider Einstein gravity with a cosmological constant in $n$ bulk dimensions:
\begin{equation}
\label{eq:action}
    S = \int d^n x\,\sqrt{-g}\,\left(R - 2\Lambda\right) \;+\; S_{\rm bdy},
\end{equation}
where $S_{\rm bdy}$ contains the Gibbons--Hawking term and local counterterms appropriate for a well-defined, finite action.
We also use units such that $16\pi G_5=1$.
Varying the bulk action yields
\begin{equation}
\label{eq:Einstein_eq}
R_{\mu\nu} - \frac12 R g_{\mu\nu} + \Lambda g_{\mu\nu} = 0
\Longleftrightarrow
R_{\mu\nu} = \frac{2}{n-2}\Lambda\, g_{\mu\nu}.
\end{equation}

Let $G$ be a three-dimensional Lie group acting simply transitively on spatial slices.
A left-invariant coframe $\{\theta^i\}$ satisfies
\begin{equation}
\label{eq:coframe_def}
d\theta^k = -\frac12 C_{ij}{}^k\,\theta^i\wedge\theta^j.
\end{equation}
For Bianchi VI$_h$ one may use the following representation:
\begin{equation}
\label{eq:BVIh_oneforms}
\theta^1 = e^{x^3}\,dx^1,\qquad
\theta^2 = e^{h x^3}\,dx^2,\qquad
\theta^3 = dx^3,
\end{equation}
with real parameter $h$.
These forms encode the spatial homogeneity and allow the bulk equations to reduce to radial ODEs.

We take a static metric of the form
\begin{equation}
\label{eq:metric_ansatz}
g = -G_t(r)\,dt^2 + G_r(r)\,dr^2 + \sum_{i=1}^3 G_i(r)\,\theta^i\theta^i,
\end{equation}
where $\theta^i$ are the Bianchi VI$_h$ one-forms \eqref{eq:BVIh_oneforms}.
Equation \eqref{eq:metric_ansatz} is homogeneous but not isotropic, so there are translational symmetries along the directions of \eqref{eq:BVIh_oneforms}.
A regular nonextremal horizon at $r=r_h$ is characterized by $G_t(r_h)=0$, with $G_t$ vanishing linearly in an appropriate radial gauge, while $G_r$ has the corresponding simple pole and $G_i(r_h)$ are finite and positive.

\subsection{Equations of motion}
Since \eqref{eq:metric_ansatz} is a cohomogeneity-one metric, substituting it into $R_{\mu\nu}=\frac{2}{n-2}\Lambda g_{\mu\nu}$ yields a coupled system of ordinary differential equations for $G_t,G_r,G_i$.
Following \cite{Hassaine:2015ifa}, we set $G_i$ and $G_r$ as follows:
\begin{equation}
\label{eq:cohomo_ansatz}
\begin{aligned}
G_t &= b_t r^{2 a_0} f(r),\\
G_r &= \frac{b_r}{r^2 f(r)},\\
G_1 &= b_1 r^{2 a_1},\\
G_2 &= b_2 r^{2 a_2},\\
G_3 &= b_3 r^{2 a_3}.
\end{aligned}
\end{equation}
The constants $b_t,b_1,b_2$ can be set to $1$ by coordinate redefinitions and a rescaling of $t$.
The normalization of $b_r$ may then be fixed so that $\lim_{r\to\infty} f(r)=1$.
Evaluating the trace-reversed Einstein equations yields five equations:
\begin{widetext}
\begin{equation}
\label{eq:cohomo_ansatz_EE_equations}
\begin{aligned}
f(r)\Big(3 r \big[(3 a_0+a_1+a_2+a_3+1) f'(r)+r f''(r)\big]
+6 a_0 (a_0+a_1+a_2+a_3) f(r)+4 \Lambda\Big)&=0, \\
\frac{1}{f(r)}\Big(3 r \big[(3 a_0+a_1+a_2+a_3+1) f'(r)+r f''(r)\big]
+6\left(a_0^2+a_1^2+a_2^2+a_3^2\right) f(r)+4 \Lambda\Big)&=0, \\
r^{2 a_3}\Big(3 a_1 \big[(a_0+a_1+a_2+a_3) f(r)+r f'(r)\big]+2 \Lambda\Big)
+\frac{3 (h+1)}{b_3}&=0, \\
r^{2 a_3}\Big(3 a_2 \big[(a_0+a_1+a_2+a_3) f(r)+r f'(r)\big]+2 \Lambda\Big)
+\frac{3 h (h+1)}{b_3}&=0, \\
r^{2 a_3}\Big(3 a_3 \big[(a_0+a_1+a_2+a_3) f(r)+r f'(r)\big]+2 \Lambda\Big)
+\frac{3 \left(h^2+1\right)}{b_3}&=0.
\end{aligned}
\end{equation}
\end{widetext}

\section{New Bianchi VI\texorpdfstring{\textsubscript{h}}{h} vacuum black brane solution}
\label{sec:solution}

In this section we discuss the derivation of the black brane solution.
Starting with \eqref{eq:cohomo_ansatz} and \eqref{eq:cohomo_ansatz_EE_equations}, we derive a solution.
First, we set $a_3 = 0$ so that the last equation of \eqref{eq:cohomo_ansatz_EE_equations} implies
\begin{equation}
\label{eq:b3_sol}
b_3 = -\frac{3 (1 + h^2)}{2 \Lambda}\,.
\end{equation}
We then solve for $f$ using the third (or fourth) equation; these also imply the following relations among the remaining parameters $a_{0,1,2}$ (together with the other unused equations):
\begin{equation}
\label{eq:a_params}
\begin{aligned}
a_0 &= \frac{1+h^2}{1-h}\,a_2,\\
a_1 &= -h\,a_2\,.
\end{aligned}
\end{equation}
Via a coordinate transformation we can set $a_2 = 1$.
This leads to the final form of our metric:
\begin{equation}
\label{eq:BVIh_metric}
\begin{split}
g &= - r^{2(1+h^2)/(1-h)} f(r)\, dt^2
+ \frac{b_r}{r^2 f(r)}\, dr^2 + r^{-2h} \left(\theta^1\right)^2\\
&\quad + r^{2} \left(\theta^2\right)^2
+ \frac{3 (1 + h^2)}{-2 \Lambda}\, \left(\theta^3\right)^2,\\
b_r &= \frac{(-\Lambda)\,(h-1)^2}{3(1+h^2)(h^2-h+1)}\,.
\end{split}
\end{equation}
The function $f$ in these final coordinates is
\begin{equation}
\label{eq:BVIh_f}
\begin{aligned}
f(r)&=
1-\left(\frac{r_h}{r}\right)^{\Delta},\\
\Delta(h)&=\frac{2(h^2-h+1)}{1-h}\,.
\end{aligned}
\end{equation}
We have scaled $t$ and $f$ so that $f\to 1$ as $r\to\infty$.
Setting $h = -1$, we recover solutions found in \cite{Hassaine:2015ifa,Cadeau:2000tj}.
Note that $\Lambda$ must be strictly negative ($\Lambda < 0$) for the metric to be Lorentzian
\footnote{One could allow $\Lambda > 0$ by performing a Wick rotation and the appropriate transformation of the radial coordinate \eqref{eq:dr_rho}. This would lead to a gravitational soliton, which could be a topic of future work.}.
Here $r_h$ is the horizon radius, defined by spacetime points with $r = r_h$ and $g(\pd_t, \pd_t) = 0$ or equivalently $f(r_h) = 0$.
Note that $b_3$ is related to the cosmological constant as $b_3 = 3 (1 + h^2)/(-2 \Lambda)$, linking the curvature of the transverse Bianchi VI$_h$ directions to the full spacetime.

\subsection{Values of \texorpdfstring{$h$}{h}}
\label{sec:values_of_h}
The value $h = 1$ yields an ill-defined expression for some components of \eqref{eq:BVIh_metric}, coinciding with a degeneration of the frame algebra.
Nonetheless, we can find a valid metric for $h=1$; see \eqref{eq:BVI1_metric}.
For $h = 0$, the frame algebra degenerates to Bianchi III.
For the special case $h = -1$, we obtain a metric with Solv geometry (Bianchi VI$_h$) as in \cite{Hassaine:2015ifa, Cadeau:2000tj}.

If $h < 1$ then $\Delta\geq 2$.
In particular, for $h < 1$ one has $f(r) > 0$ for all $r > r_h$.
Thus we can identify the exterior of the brane with the region $r>r_h$.

On the other hand, if $h>1$ then $f(r) > 0$ for all $r<r_h$.
In this case the exterior region is $r<r_h$; however, due to the $r$-dependence of other metric components (which vanish as $r\to 0$), it is clearer to introduce an inverted coordinate $r = 1/\rho$, so that
\begin{equation}
\label{eq:dr_rho}
dr = -\frac{d\rho}{\rho^2}\,.
\end{equation}
In contrast to \eqref{eq:BVIh_metric}, the resulting metric is
\begin{equation}
\label{eq:BVIh_metric_rho}
\begin{split}
g &= - \rho^{-2(1+h^2)/(1-h)} f(\rho^{-1})\, dt^2
+ \frac{b_r}{\rho^2 f(\rho^{-1})}\, d\rho^2 \\
&\quad + \rho^{2h} \left(\theta^1\right)^2
+ \rho^{-2} \left(\theta^2\right)^2
+ \frac{3 (1 + h^2)}{-2 \Lambda}\, \left(\theta^3\right)^2\,,
\end{split}
\end{equation}
It is then apparent that if $h>1$ then the resulting black brane is causally well-behaved.
Relative to the $h<1$ case, the brane is ``inside out'', but this is simply a consequence of our parametrization of the Bianchi VI$_h$ algebras.
Equivalently, one can replace $\Delta$ with $|\Delta|$ and $a_0$ with $|a_0|$ in \eqref{eq:BVIh_f} and substitute $r \to r^{-1}$ for the non-bulk transverse directions to obtain a uniform presentation for all $h\neq 1$, \eqref{eq:BVIh_metric_allh}.
\begin{equation}
\label{eq:BVIh_metric_allh}
\begin{split}
g &= - r^{2(1+h^2)/|1-h|} f(r)\, dt^2
+ \frac{b_r}{r^2 f(r)}\, dr^2 \\
&\quad + r^{-2h\sign(1-h)} \left(\theta^1\right)^2
+ r^{2\sign(1-h)} \left(\theta^2\right)^2\\
&\quad + \frac{3 (1 + h^2)}{-2 \Lambda}\, \left(\theta^3\right)^2,\\
b_r &= \frac{(-\Lambda)\,(h-1)^2}{3(1+h^2)(h^2-h+1)}\,.
\end{split}
\end{equation}
The function $f$ in these coordinates is
\begin{equation}
\label{eq:BVIh_f_allh}
\begin{aligned}
f(r)&=
1-\left(\frac{r_h}{r}\right)^{|\Delta|},\\
|\Delta|&=\frac{2(h^2-h+1)}{|1-h|}\,.
\end{aligned}
\end{equation}

\subsection{Special case: Bianchi V (\texorpdfstring{$h = 1$}{h = 1})}

When $h=1$ the ansatz \eqref{eq:cohomo_ansatz} breaks down if $a_1 \neq 0$ or $a_2 \neq 0$.
In that case, we find a solution by setting $a_1 = a_2 = a_3 = 0$.
Solving again for $f$ and the remaining coefficients we get
\begin{equation}
\label{eq:BVI1_metric}
\begin{aligned}
g &= - f(r)\, dt^2
+ \frac{b_r}{f(r)}\, dr^2
+ e^{2x^3} \left(dx^1\right)^2 \\
&\quad + e^{2x^3} \left(dx^2\right)^2
+ \frac{3}{-\Lambda}\, \left(dx^3\right)^2,\\
b_r &= \frac{3}{-2 \Lambda}\,,
\end{aligned}
\end{equation}
where the blackening factor is
\begin{equation}
\label{eq:BVI1_f}
f(r) = (r-r_h) (r - r_m)\,.
\end{equation}
Without loss of generality, $r_m < r_h$ is a free integration constant, which is the second root of the blackening factor.
As in the general case, $\Lambda$ must be strictly negative for the metric to be Lorentzian.
The discontinuity of the metric with respect to variations in $h$ at $h=1$ makes the analysis more delicate, so the solution \eqref{eq:BVI1_metric} is not necessarily related to \eqref{eq:BVIh_metric}.
Nevertheless, the geometry in the transverse directions is similar (in particular, the $g_{33}$ component is the same).
It is worth noting that for $h = 1$, the Bianchi VI$_1$ algebra degenerates to Bianchi V.
The 3D metric formed by the left-invariant Bianchi V one-forms is an Einstein metric with a negative cosmological constant.
One can also write the metric as
\begin{equation}
\label{eq:BV_uni_metric}
\begin{aligned}
g &= - r^2 f(r)\, dt^2
+ \frac{b_r}{r^2 f(r)}\, dr^2
+ r^2 e^{2x^3} \left(dx^1\right)^2 \\
&\quad + r^2 e^{2x^3} \left(dx^2\right)^2
+ b_3 r^2 \left(dx^3\right)^2,\\
b_r &= \frac{6}{-\Lambda}\,,
\end{aligned}
\end{equation}
where the blackening factor is
\begin{equation}
\label{eq:BV_f}
f(r) = \frac{(r-r_h) (r+r_h)\left(b_3 (-\Lambda) \left(r^2+r_h^2\right)-6\right)}{b_3 (-\Lambda) r^4}\,.
\end{equation}
This metric is the well-known hyperbolic AdS black brane \cite{Birmingham:1998nr,Emparan:1999gf}, with $\Lambda$ set to the appropriate value $\Lambda = -6/L^2$.

\subsection{Curvature}

We characterize curvature singularities using scalar invariants. The ones we use are
\begin{equation}
\label{eq:inv_defs}
R := g^{\mu\nu}R_{\mu\nu},\quad
R_{\mu\nu}R^{\mu\nu},\quad
K := R_{\mu\nu\rho\sigma}R^{\mu\nu\rho\sigma}\,.
\end{equation}
For the metrics \eqref{eq:BVIh_metric}--\eqref{eq:BVIh_f} (which solve the vacuum Einstein equations in $n=5$ dimensions), the spacetime is Einstein:
\begin{equation}
\label{eq:Einstein_tensor}
R_{\mu\nu}=\frac{2}{3}\Lambda\, g_{\mu\nu}\,.
\end{equation}
Consequently the Ricci invariants are constant everywhere:
\begin{equation}
\label{eq:Ricci_invariants}
R=\frac{10}{3}\Lambda,\qquad
R_{\mu\nu}R^{\mu\nu}=\frac{20}{9}\Lambda^2\,.
\end{equation}
The Kretschmann scalar may still signal the presence of a curvature singularity.
A direct computation of the Kretschmann scalar yields a compact result:
\begin{equation}
\label{eq:BVIh_Kretschmann}
\begin{split}
K(r) &= \frac{28}{9}\Lambda^2
+ \frac{16\,\Lambda^2\,h^2(h-1)^2}{9\,(h^2-h+1)\,(h^2+1)^2} \times \\
&\quad \left[ \left(\frac{r_h}{r}\right)^{2|\Delta|} - \left(\frac{r_h}{r}\right)^{|\Delta|} \right].
\end{split}
\end{equation}
If $h<1$ then the asymptotic region is $r\to\infty$, and one has
\begin{equation}
\label{eq:K_limit}
K(r)\xrightarrow[r\to\infty]{}\frac{28}{9}\Lambda^2.
\end{equation}
Also, as $r\to 0$ the leading behavior is
\begin{equation}
\label{eq:K_singularity}
K(r)\sim
\frac{16\,\Lambda^2\,h^2(h-1)^2}{9\,(h^2-h+1)\,(h^2+1)^2}\,
\left(\frac{r_h}{r}\right)^{2|\Delta|}\,,
\end{equation}
so $r=0$ is a genuine curvature singularity unless the prefactor vanishes.
Nevertheless, if $h=0$ the $r$-dependent term vanishes:
\begin{equation}
\label{eq:K_h0}
K^{h=0}(r)\equiv \frac{28}{9}\Lambda^2.
\end{equation}
Nevertheless, these spacetimes are not of constant curvature (unlike AdS$_5$).
For a constant-curvature spacetime with cosmological constant $\Lambda$, one needs
\begin{equation}
\label{eq:K_constant}
K(r) = K_{\rm CONST.} \equiv \frac{10}{9}\Lambda^2\,.
\end{equation}
Since our solutions instead asymptote to $K\to \frac{28}{9}\Lambda^2$, they differ from those that are asymptotically locally AdS.

Equation \eqref{eq:BVI1_metric} has constant $R$ and $R_{\mu\nu}R^{\mu\nu}$ curvature invariants.
Strikingly, the Kretschmann invariant is constant, $K = \frac{28}{9}\Lambda^2$, and is identical to the $h = 0$ case.
In both cases, there is an event horizon that causally separates two regions of spacetime.

\subsection{Constant-Curvature Cases}
We have seen that there is no curvature singularity for \eqref{eq:BVI1_metric}.
For special values of $h$, selected curvature invariants were found to be finite, which hinted at the absence of curvature singularities.
Nevertheless, here we will examine the global geometry of these special geometries, starting with the $h = 1$ case.
We can shift the $r$ coordinate to $\rho$ such that $r = \rho + (r_h + r_m)/2$.
In terms of $\rho$, \eqref{eq:BVI1_f} becomes $f = (\rho - \rho_0) (\rho + \rho_0) = \rho^2 - \rho_0^2$ where $\rho_0 = (r_h - r_m)/2$.
In these coordinates, the metric becomes
\begin{equation}
\label{eq:BVI1_metric_hyper}
\begin{split}
g &= - (\rho^2 - \rho_0^2)\, dt^2
+ \frac{3/(-2 \Lambda)}{\rho^2 - \rho_0^2}\, d\rho^2 \\
&\quad + e^{2x^3} \left(dx^1\right)^2
+ e^{2x^3} \left(dx^2\right)^2
+ \frac{3}{-\Lambda}\, \left(dx^3\right)^2.
\end{split}
\end{equation}
This metric is a Rindler wedge of AdS$_2$ times hyperbolic space $H^3$, with the AdS radius determined by $\Lambda$.
One can see this via the coordinate transformation $t \rightarrow t'/\rho_0$ and $\rho \rightarrow \rho_0 \cosh( r' )$.
This is similar to the case where $h=0$, for which \eqref{eq:BVIh_metric} becomes
\begin{equation}
\label{eq:BVI0_metric}
\begin{split}
g &= -\left(r^2 - r_h^2\right) dt^2
+ \frac{-\Lambda}{3\left(r^2 - r_h^2\right)}\, dr^2
+ r^{2} \left(dx^2\right)^2 \\
&\quad + e^{2x^3} \left(dx^1\right)^2
+ \frac{3}{-2 \Lambda}\, \left(dx^3\right)^2.
\end{split}
\end{equation}
This spacetime is a direct product of a BTZ black hole \cite{Banados:1992gq} and $H^2$.
The near-horizon geometry of supersymmetric spacetimes found in \cite{Klemm:2000nj,Karndumri:2022xmp} is equivalent.
From this, it is natural to suppose that the $h\neq 0$ solutions are near-horizon geometries of more general spacetimes.
There is no curvature singularity, but it is well known that there is a causal singularity for the nonspinning BTZ black hole \cite{Banados:1992gq}.

\subsection{Thermodynamics}
\label{sec:thermo}

In this section we compute the temperature of the Bianchi VI$_h$
black branes. Throughout we work in $n=5$ bulk dimensions with action normalization
$16\pi G_5=1$.
For $h\neq 1$ we use the metric \eqref{eq:BVIh_metric_allh}--%
\eqref{eq:BVIh_f_allh}.
The special case, Bianchi V ($h=1$), \eqref{eq:BVI1_metric} is treated separately.

For convenience we rewrite the $h\neq 1$ metric in the standard static form
\begin{equation}
\label{eq:static_form}
ds^2 = -A(r)\,dt^2 + B(r)\,dr^2 + ds^2_{\rm hor}(r),
\end{equation}
where from \eqref{eq:BVIh_metric}--\eqref{eq:BVIh_f}
\begin{equation}
\label{eq:AB_defs}
\begin{aligned}
A(r)&= r^{2a_0}\,f(r),\qquad
B(r)=\frac{b_r}{r^2 f(r)},\\
f(r)&=1-\left(\frac{r_h}{r}\right)^{\Delta}.
\end{aligned}
\end{equation}
The exponents and constants are
\begin{equation}
\label{eq:thermo_params}
\begin{aligned}
a_0&=\frac{1+h^2}{|1-h|},\quad
\Delta=\frac{2(h^2-h+1)}{|1-h|},\\
b_r&=\frac{(-\Lambda)(h-1)^2}{3(1+h^2)(h^2-h+1)},\quad
b_3=\frac{3(1+h^2)}{-2\Lambda}.
\end{aligned}
\end{equation}

\subsection{Temperature}
The Hawking temperature is fixed by regularity of imaginary time at the
horizon. For a static metric $ds^2=-A(r)\,dt^2+B(r)\,dr^2+\cdots$ with a simple zero of
$A(r)$ at $r=r_h$, one may use
\begin{equation}
\label{eq:T_general}
T=\frac{1}{4\pi}\left.\frac{A'(r)}{\sqrt{A(r)B(r)}}\right|_{r=r_h}.
\end{equation}
Therefore
\begin{equation}
\label{eq:BVIh_temperature_clean}
\begin{split}
T
&=\frac{1}{4\pi}\frac{r_h^{2a_0} f'(r_h)}{\sqrt{b_r}\,r_h^{a_0-1}}
=\frac{\Delta}{4\pi\sqrt{b_r}}\,r_h^{a_0}\\
&=\frac{1}{4\pi\sqrt{b_r}}\, \frac{2(h^2-h+1)}{|1-h|}\; r_h^{\frac{1+h^2}{|1-h|}}.
\end{split}
\end{equation}

\subsection{Entropy}
\label{sec:thermo_entropy}

We now compute the (Bekenstein--Hawking) entropy from the horizon area:
\begin{equation}
\label{eq:entropy_def}
S=\frac{A_h}{4G_5}=4\pi A_h.
\end{equation}
For the solutions \eqref{eq:BVIh_metric_allh}-- \eqref{eq:BVIh_f_allh}, the metric on a constant-$(t,r)$ slice is diagonal in the left-invariant coframe $\{\theta^i\}$. It is therefore natural to express the horizon area in terms of the invariant horizon volume
\begin{equation}
\label{eq:volume_def}
V_3 \;:=\;\int_{\Sigma_3}\theta^1\wedge\theta^2\wedge\theta^3 .
\end{equation}
At $r=r_h$ the induced metric on the spatial cross section of the horizon is
\begin{equation}
\label{eq:hor_induced}
ds^2_{\rm hor}\big|_{r_h} =
r_h^{-2h\sign(1-h)}(\theta^1)^2 + r_h^{2\sign(1-h)}(\theta^2)^2 + b_3(\theta^3)^2 .
\end{equation}
The horizon area is therefore \footnote{Note that for $h > 1$ in \eqref{eq:BVIh_metric_allh}, the $r$ coordinate ``flips.''}
\begin{equation}
\label{eq:area_formula}
A_h = V_3\,\sqrt{b_3}\; r_h^{|1-h|}.
\end{equation}
The entropy and entropy density (per unit invariant volume) become
\begin{equation}
\label{eq:BVIh_entropy_clean}
S = 4\pi\,V_3\,\sqrt{b_3}\; r_h^{\,|1-h|}, \quad
s := \frac{S}{V_3} = 4\pi\,\sqrt{b_3}\; r_h^{\,|1-h|}.
\end{equation}
Thus the entropy scales with the horizon radius as
\begin{equation}
\label{eq:s_scaling}
s \propto r_h^{\,|1-h|}.
\end{equation}
Combining this with the temperature scaling
$T\propto r_h^{|a_0|}$ from \eqref{eq:BVIh_temperature_clean},
one obtains the power-law relation
\begin{equation}
\label{eq:s_T_rel}
s \propto T^{\left|\frac{1-h}{a_0}\right|}.
\end{equation}
We can compare the entropy with hyperscaling relations $s \sim T^{(d-\theta)/z}$ \cite{Huijse:2011ef,Gouteraux:2011ce}.
With $z = |a_0| = (1+h^2)/|1-h|$ and $d = 3$ (the number of non-bulk spatial dimensions), we may identify $\theta = h + 2$ if $h < 1$ or $\theta = 4 - h$ if $h > 1$.
The parameter $h$ then acts like an effective dimensional shift.
For the special value $h=0$ or $h=2$, $\theta = 2 = d - 1$, which is characteristic of a Landau Fermi liquid \cite{Pedraza:2018eey,Huijse:2011ef}.

For $h=1$ we have $A(r)=f(r)$ and $B(r)=b_r/f(r)$.
Euclidean regularity at the Rindler horizon requires the temperature to be
\begin{equation}
\label{eq:T_h1}
T_{h=1}=\frac{1}{4\pi}\frac{f'(r_h)}{\sqrt{b_r}}
=\frac{r_h-r_m}{4\pi\sqrt{b_r}}
=\frac{r_h-r_m}{4\pi}\sqrt{\frac{-2\Lambda}{3}}\,,
\end{equation}
which is positive for $r_m<r_h$.

\subsection{Specific Heat Capacity}

The specific heat capacity at fixed volume is
\begin{equation}
\label{eq:heat_capacity_def}
c_V = T\frac{ds}{dT}.
\end{equation}
The unified explicit expression valid for both $h<1$ and $h>1$ becomes
\begin{equation}
\label{eq:specific_heat}
c_V \propto \left|\frac{1-h}{a_0}\right|\, T^{\left|\frac{1-h}{a_0}\right|}.
\end{equation}
Equation \eqref{eq:specific_heat} also works in the case $h = 1$, where $c_V$ vanishes.

\section{The \texorpdfstring{$\Lambda=0$}{Lambda=0} case and hyperscaling violation}
\label{sec:lambda_zero}

In addition to the solutions supported by a negative cosmological constant, the system admits a solution in the limit $\Lambda = 0$. This corresponds to a Ricci-flat vacuum black brane with Bianchi VI$_h$ horizon topology.

Hassa\"ine et al.\ \cite{Hassaine:2015ifa} previously investigated homogeneous solutions in this limit. They successfully identified a vacuum solution for the Nil geometry (Bianchi II), which exhibits hyperscaling violation. However, they noted that analogous solutions for the Solv geometry (Bianchi VI$_{-1}$) were unattainable in the vacuum limit. Our result generalizes this to the generic Bianchi VI$_h$ family, showing that solutions exist for $h \neq -1$ but degenerate at the Solv point.

Starting from the generic ansatz and setting $\Lambda=0$ in the Einstein equations, we find that the metric takes a form characteristic of hyperscaling-violating geometries. Incorporating the conformal factor $\Omega(r) = r^{-2\theta/3}$, the metric is:
\begin{equation}
\label{eq:BVIh_vacuum_metric}
\begin{split}
ds^2 &= r^{-\frac{2\theta}{3}} \Bigg[
-r^{2z} f(r)\, dt^2 + \frac{b_r}{r^2 f(r)}\, dr^2 \\
     & + r^{-2h\sign(1-h)} (\theta^1)^2 + r^{2 \sign(1-h)} (\theta^2)^2\\
     &+ b_3 (\theta^3)^2
\Bigg].
\end{split}
\end{equation}
Here $b_3$ is an integration constant. The parameters determined by the equations of motion are:
\begin{equation}
\label{eq:vac_params}
z = \frac{1+h^2}{|1-h|}, \quad \theta = \frac{3(1+h^2)}{|1-h|}, \quad b_r = b_3\left(\frac{h+1}{h-1}\right)^2.
\end{equation}
Notably, the dynamical critical exponent $z$ and the hyperscaling violation exponent $\theta$ are fixed by the anisotropy parameter $h$, satisfying the relation $\theta = 3z$.
Also note that we used the same trick as in Section~\ref{sec:values_of_h} for $h>1$, to keep the exterior of the black brane at $r > r_h$.

The blackening factor $f(r)$ is given by:
\begin{equation}
\label{eq:vacuum_f}
f(r) = -1 + \left(\frac{r}{r_h}\right)^\gamma, \qquad \gamma = \frac{(1+h)^2}{|1-h|}.
\end{equation}
For the solution to represent a black hole with a horizon at $r=r_h$, we require $\gamma > 0$, which implies $h < 1$. Unlike the asymptotically AdS ($\Lambda < 0$) case where $f \to 1$, here $f(r)$ grows polynomially at large $r$, a common feature of vacuum scaling solutions.

The radial coefficient $b_r$ vanishes as $h \to -1$. This indicates that the metric degenerates exactly at the Solv point. This analytical result explains the observation in \cite{Hassaine:2015ifa} regarding the absence of Solv vacuum solutions: the Solv geometry ($h=-1$) specifically requires $\Lambda \neq 0$ (or other matter support) to remain nondegenerate.


\subsection{Thermodynamics}
The temperature of this Ricci-flat solution is computed using the surface gravity. Using the derived metric components and assuming $|h| \neq 1$, the temperature simplifies to:
\begin{equation}
\label{eq:vacuum_temp}
T = \frac{|1+h|}{4\pi}\, r_h^{\frac{1+h^2}{|1-h|}} = \frac{|1+h|}{4\pi}\, r_h^{z}.
\end{equation}
This confirms that the Bianchi VI$_h$ anisotropy supports a thermodynamic black brane solution in vacuum, with the temperature scaling governed by the effective critical exponent $z$.
Similarly to Section~\ref{sec:thermo} we can calculate the entropy density as
\begin{equation}
\label{eq:vac_entropy}
s =  \frac{1}{4}\sqrt{b_3}\, r_h^{\left(|1-h|-\theta\right)} = \frac{1}{4}\sqrt{b_3}\, r_h^{|1-h|-\frac{3(1+h^2)}{|1-h|}}\,.
\end{equation}
This implies that $s \propto T^{(|1-h|-\theta)/z}$.
Comparing with a hyperscaling-violating relation, an effective number of spatial dimensions is $d_{\rm eff} = |h-1|$.
For the case $h = 1$, this effective number of dimensions vanishes.

\subsection{Special case: Bianchi V (\texorpdfstring{$h = 1$}{h = 1})}
\label{sec:vacuum_bianchiV_h0}

With $h=1$ the frame algebra reduces to Bianchi V.
To find a solution we choose the hyperscaling exponent and time-warp exponent as
\begin{equation}
\label{eq:BV_exponents}
\theta = 3a_0,\qquad a_0=1\quad(\Rightarrow\ \theta=3),
\end{equation}
and the field equations admit the Ricci-flat black brane metric
\begin{equation}
\label{eq:BV_vacuum_metric_h0}
\begin{split}
ds^2
&= r^{-2}\Bigg[
- r^{2} f(r)\,dt^2
+ \frac{b_3}{r^{2} f(r)}\,dr^2\\
&\quad + (\theta^1)^2+(\theta^2)^2
+ b_3(\theta^3)^2
\Bigg],
\end{split}
\end{equation}
where $b_3$ is a positive constant and the blackening factor is
\begin{equation}
\label{eq:BV_vacuum_f_h0}
f(r)=\left(\frac{r}{r_h}\right)-1.
\end{equation}
In particular, $f(r_h)=0$ and the exterior region is $r>r_h$, where $f(r)>0$.
This solution has $z=1$ and $\theta=3$ in the sense of the overall conformal
prefactor $r^{-2\theta/3}=r^{-2}$.

\section{Conclusion}
\label{sec:conclusion}

We have constructed a new family of homogeneous vacuum black brane solutions of five-dimensional Einstein gravity with Bianchi VI$_h$ symmetry.
The geometry is controlled by a continuous anisotropy parameter $h$, interpolating between previously known special cases such as the Solv solution at $h=-1$ and degenerate branches at $h=0$ and $h=1$.
The Einstein equations reduce to a cohomogeneity-one ODE system, allowing for an explicit analytic presentation of the metric, thermodynamics, and curvature invariants.

Although the spacetime is Einstein with a negative cosmological constant, it is not asymptotically locally AdS.
Instead, it exhibits anisotropic power-law behavior and approaches a non-constant-curvature geometry at large radius (for $h<1$).
In special cases the metric becomes a direct product involving AdS$_2$ factors, and more generally the structure strongly suggests that the Bianchi VI$_h$ geometries should be interpreted as infrared scaling solutions.
In particular, they are natural candidates for near-horizon geometries of more general spacetimes, potentially arising as the deep-interior limit of asymptotically locally AdS black branes.

From a holographic standpoint, this raises an immediate and concrete question: does there exist a domain-wall solution that interpolates between AdS$_5$ in the ultraviolet and a Bianchi VI$_h$ geometry in the infrared?
If so, the present solutions would represent homogeneous anisotropic IR fixed points of RG flows driven by symmetry-breaking deformations.
Because the construction works already in pure Einstein gravity, any such embedding would be particularly economical.

Several extensions suggest themselves.
First, a systematic study of linear perturbations and quasinormal modes would clarify dynamical stability and transport properties.
Second, one may attempt to construct explicit interpolating geometries, either numerically or with suitable matter couplings (e.g.\ Maxwell or Einstein--Maxwell--dilaton sectors) that naturally generate flows between AdS and anisotropic scaling regimes.
Third, it would be valuable to understand the classification problem more globally: whether different Bianchi types can be connected through smooth families of solutions or whether the parameter space decomposes into disconnected branches.

At minimum, we have shown that the Bianchi VI$_h$ family provides a continuous deformation that includes the known Solv sector and extends it into genuinely new anisotropic regimes.
Viewed geometrically, these metrics enlarge the catalogue of explicit homogeneous Einstein spacetimes.
It would be interesting to understand how, in general, one can compactify the horizon spacetimes, which could be related to potential field theory period systems that are symmetric with respect to finite group actions.

Furthermore, we have demonstrated that the Bianchi VI$_h$ ansatz is robust enough to support solutions in the vanishing cosmological constant limit. We derived an explicit Ricci-flat metric family characterized by the same anisotropy parameter $h$. These vacuum solutions exhibit polynomial growth in the blackening factor and a simple power-law temperature dependence, $T \propto r_h^z$, suggesting they may have some use for some holographic systems.

\begin{acknowledgements}
The author would like to thank Guilherme Sadovski for valuable feedback and discussion.
The author was provided with funding by the National Institute of Technology, Oyama College.
\end{acknowledgements}

\appendix
\section{General $\Lambda < 0$ Bianchi II/IV$_{-1}$ Black Branes}
\label{app:hassaine_generic}

For comparison with our Bianchi VI$_h$ family, we collect here the generic (i.e.\ with arbitrary $\Lambda<0$ and without fixing normalizations) form of the homogeneous vacuum Einstein black branes discussed by Hassa\"ine and collaborators \cite{Hassaine:2015ifa} (see also Cadeau--Woolgar \cite{Cadeau:2000tj} for closely related Thurston-horizon solutions).
In each case the key mechanism is the same as in the main text: once the spatial slices are written in a left-invariant coframe, the Einstein equations reduce to a cohomogeneity-one ODE for the blackening factor.

\subsection{Nil horizon (Bianchi II)}
\label{app:nil}

Let $\{\theta^i\}$ be the standard left-invariant one-forms on the Heisenberg (Nil) group,
\begin{equation}
\label{eq:Nil_forms}
\theta^1=dx^1,\qquad \theta^2=dx^2,\qquad \theta^3=dx^3-x^1\,dx^2,
\end{equation}
so that $d\theta^3=-\theta^1\wedge\theta^2$ and $d\theta^1=d\theta^2=0$.
A generic Einstein Nil-horizon black brane of the type studied in
\cite{Hassaine:2015ifa} may be written as
\begin{equation}
\label{eq:Nil_generic_metric}
\begin{aligned}
ds^2
&= - r^{3}\,f(r)\,dt^2
+ \frac{dr^2}{r^{2} f(r)}\\
&\quad + r^{2}\Big[(\theta^1)^2+(\theta^2)^2\Big]
+ \Big(-\frac{4\Lambda}{9}\Big)\,r^{4}\,(\theta^3)^2,
\end{aligned}
\end{equation}
with blackening factor determined by the Einstein equations as
\begin{equation}
\label{eq:Nil_generic_f}
f(r)= -\frac{8\Lambda}{99}\left[1-\left(\frac{r_h}{r}\right)^{11/2}\right].
\end{equation}
Since $\Lambda<0$, the prefactor $-\frac{8\Lambda}{99}$ is positive and can be absorbed into a rescaling of $t$ (equivalently, one can redefine $f$ so that $f\to 1$ asymptotically).
In particular, the normalization chosen in \cite{Hassaine:2015ifa} corresponds to fixing $\Lambda=-99/8$, for which \eqref{eq:Nil_generic_f} becomes $f(r)=1-(r_h/r)^{11/2}$ and the coefficient of $(\theta^3)^2$ becomes $11/2$.

\subsection{Solv horizon (Bianchi VI$_{-1}$)}
\label{app:solv}

Now take the Solv (equivalently Bianchi VI$_{-1}$) left-invariant one-forms
\begin{equation}
\label{eq:Solv_forms}
\theta^1=e^{x^3}dx^1,\qquad
\theta^2=e^{-x^3}dx^2,\qquad
\theta^3=dx^3,
\end{equation}
for which $d\theta^1=\theta^1\wedge\theta^3$ and $d\theta^2=-\theta^2\wedge\theta^3$.
A corresponding generic Einstein black brane can be written as
\begin{equation}
\label{eq:Solv_generic_metric}
\begin{split}
ds^2
&= - r^{2}\,f(r)\,dt^2
+ \frac{dr^2}{r^{2} f(r)}\\
&\quad + r^{2}\Big[(\theta^1)^2+(\theta^2)^2\Big]
+ \Big(-\frac{3}{\Lambda}\Big)\,(\theta^3)^2,
\end{split}
\end{equation}
with
\begin{equation}
\label{eq:Solv_generic_f}
f(r)= -\frac{2\Lambda}{9}\left[1-\left(\frac{r_h}{r}\right)^3\right].
\end{equation}
Again $-\frac{2\Lambda}{9}>0$ for $\Lambda<0$ and may be absorbed into a rescaling of $t$.
The parameter choice $\Lambda=-9/2$ used in \cite{Hassaine:2015ifa} yields the particularly simple form $f(r)=1-(r_h/r)^3$ together with the constant coefficient $(\theta^3)^2$ term equal to $2/3$, reproducing the familiar Solv-horizon sector also discussed in \cite{Cadeau:2000tj}.

\bibliographystyle{unsrturl}
\bibliography{references}

\end{document}